\documentclass[preprint,12pt]{elsarticle}
\usepackage{color}

\usepackage{graphics}

\usepackage{amssymb}

\usepackage{lineno}

\journal{NIM A }

\begin{document}

\begin{frontmatter}



\title{Validation of a new background discrimination method for the TACTIC TeV $\gamma$-ray telescope with Markarian 421 data}


\author[lab1]{Mradul Sharma\footnote{\small mradul@barc.gov.in}}
\author[lab2]{J. Nayak}
\author[lab1]{M.K. Koul}
\author[lab2]{S. Bose}
\author[lab1]{Abhas Mitra}
\author[lab1]{V.K.Dhar}
\author[lab1]{A.K.Tickoo}
\author[lab1]{R. Koul}
\address[lab1]{Astrophysical Sciences Division, Bhabha Atomic Research Centre, Mumbai, India}
\address[lab2]{The Bayesian and Interdisciplinary Research Unit, Indian Statistical Institute,Kolkata , India}

\begin{abstract}
This paper describes the validation of a new background discrimination method based on Random Forest technique by re-analysing the Markarian 421 (Mrk 421) observations performed by the TACTIC (TeV Atmospheric Cherenkov Telescope with Imaging Camera) $\gamma$-ray telescope. The Random Forest technique is a flexible multivariate method which combines Bagging and Random Split Selection to construct a large collection of decision trees and then combines them to construct a common classifier. Markarian 421 in a high state was observed by TACTIC during December 07, 2005 - April 30, 2006 for 202 h. Previous analysis of this data led to a detection of flaring activity from the source at Energy $ >$ 1 TeV. Within this data set, a spell of 97 h revealed strong detection of a $\gamma$-ray signal with daily flux of $> 1$ Crab unit on several days. Here we re-analyze this spell as well as the data from the entire observation period with the Random Forest method.  
Application of this method led to an improvement in the signal detection strength by $\sim 26\%$ along with a $\sim 18\%$ increase in detected $\gamma$ rays compared to the conventional Dynamic Supercuts method. The resultant differential spectrum obtained is represented by a power law with an exponential cut off  $\Gamma = -2.51 \pm  0.10$ and $E_{0} = 4.71 \pm 2.20$ TeV. Such a spectrum is consistent with previously reported results and justifies the use of Random Forest method for analyzing data from atmospheric Cherenkov telescopes.
\end{abstract}


\end{frontmatter}


\section{Introduction}
\label{intro}
The active galactic nucleus (AGN) Markarian (Mrk) 421 is the first extragalactic source observed in the TeV energy range \cite{punch}. It is one 
of the blazars most widely observed by Imaging Atmospheric Cherenkov Technique (IACT). The observation of this source by various telescopes has produced a 
great wealth of scientific insights into the working of blazars in the TeV energy regime. Flux variation on varied time scales from as small as few minutes 
to as large as few years has turned this source into a laboratory for understanding the high energy emission mechanisms. The flux doubling time of as short
as $15$ minutes \cite{gaidos} as well as flux variation by more than one order of magnitude 
\cite{fossati2008} have also been observed. 
Mrk421 in the high state was observed in $2005-2006$ by various IACTs. The MAGIC telescope observed this source from April 22 to April 30, 2006 
\cite{magic2010}. The Whipple telescope \cite{horan2009} carried out the observation in April and June 2006. This source was also observed by the TACTIC 
telescope from 27 December 2005 - 30 April 2006 for  a total of $\sim 202$ h \cite{tactic2007}. Previous analysis of a  spell of $\sim 97$ h of this data 
revealed strong detection of a $\gamma$-ray signal with daily flux of  $> 1$ Crab unit at Energy $ > 1$ TeV.  
The analysis was carried out by the Dynamic Supercuts method \cite{mohanty}. Here we re-analyze the entire data set using a  
tree based classifier method named Random Forest. 
The earlier efforts to employ multivariate methods for $\gamma$/hadron segregation were initiated by Bock et. al. \cite{bock}. Later on, the effectiveness of 
tree based classifiers was demonstrated by two operational IACT systems, MAGIC \cite{magic} and HESS \cite{hess1,hess2,hess3,Bec}. In particular, the Random Forest method was employed by the MAGIC telescope \cite{hengstebeck}. 

The structure of the paper is as follows: The first section presents a short description of the TACTIC telescope. Next section describes 
the database used in training the Random Forest. The subsequent section compares the excess $\gamma$-ray events obtained by Random Forest method for different 
spells of observation with respect to the previously reported results \cite{tactic2007}. Finally, we conclude this paper by estimating the energy spectrum of Mrk 421.



\section{The TACTIC Telescope}
The TACTIC $\gamma$-ray telescope \cite{rkoul} has been in operation at Mt. Abu ($24.6^{\circ} N$, $72.7^{\circ}E, \sim 1300$ m asl) in Western India 
for the study of TeV $\gamma$-ray emissions from celestial sources.  The telescope employs a  349-pixel photomultiplier tube imaging camera with a uniform pixel resolution of 
$\sim$ $0.3^{\circ}$ and a $\sim$ $5.9^{\circ}$$\times$$5.9^{\circ}$ field-of-view to collect  atmospheric Cherenkov events generated by extensive air shower due to 
charged cosmic-rays or $\gamma$-rays. 
The  TACTIC telescope uses an F/1 type tracking light-collector of $\sim 9.5$ m$^2$ area. It consists of 34 front-face aluminum-coated  glass spherical mirrors 
of 60 cm diameter  each  with  a focal length $\sim$ 400cm. The innermost 121 pixels (11 $\times$ 11 matrix) are used for generating the event trigger based on the philosophy of 
Nearest Neighbor Pairs (NNP). 
The telescope has a pointing and tracking accuracy of better than  $\pm$3 arc-minutes. The tracking accuracy is checked on a regular 
basis with so called \textquotedblleft pointing runs\textquotedblright, where an optical star having its declination close to that of the candidate 
$\gamma$-ray source is tracked continuously for about 5 hours. 
The pointing run calibration data is then incorporated so that appropriate corrections can be applied.   
TACTIC records cosmic-ray events with a rate of $\sim$ 2.0 Hz at a typical zenith angle of $15^{\circ}$ and is operating at a $\gamma$-ray threshold energy of $\sim$ 1 TeV. 
The TACTIC readout employs charge integration over the full gating window. 
\section{Image parameters for classification}
\label{subsect:image}
\begin{figure}[!h]
\begin{center}
\includegraphics[totalheight=3.5cm]{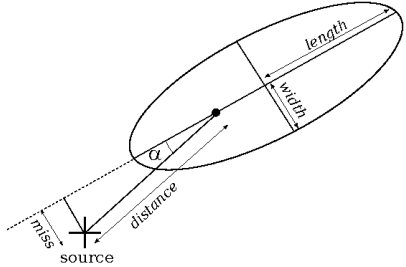}
\caption{{\it Hillas parameters}}\label{Figure:image}
\label{fig_had}
\end{center}
\end{figure}

The spatial distribution of Cherenkov photons on the image plane of the camera is known as the \textquotedblleft image\textquotedblright. The parameters used to describe the 
image are known as the Hillas parameters \cite{Hillas}. A typical spatial distribution of Cherenkov photons on the camera can be parameterized 
as an ellipse. Figure \ref{Figure:image} shows the schematic representation of Hillas parameters. The rms spread of Cherenkov light along the major/minor axis of 
image is known as the length/width of an image. The length and width parameter are a measure of the lateral and vertical development of the shower. The distance parameter represents the 
distance from the image centroid to the centre of the field of view. It is a crude measure of the core distance of the $\gamma$-ray initiated shower. The image parameter frac2 represents 
the degree of light concentration. It is defined as the ratio of the sum of the two highest pixel signal to the sum of all the signals. The size parameter represents the total number 
of photoelectrons contained in the Cherenkov image. The parameter $\alpha$ is the angle of the image and a line joining the centroid of the image to the centre of the field of view. 
It is a measure of the orientation of the shower axis.

Above image parameters were used as classifying parameters for $\gamma$/hadron segregation. 
In addition to the above parameters, a new derived parameter \textquotedblleft dens\textquotedblright, defined \cite{hengstebeck} as
\begin{equation}
\mathrm{dens}=\frac{\mathrm{log_{10}}\mathrm{(size)}} {\mathrm{length} \times \mathrm{width}}
\end{equation}  was used. 

\subsection{Monte Carlo and real data samples used for the method validation}
The TACTIC telescope observed Mrk 421 for a total of $\sim 202$ h between December $07, 2005$ - April  $30, 2006$. The source was observed for six different 
lunations labeled as Spell I - Spell VI. The observation was restricted to zenith angles $\leq 45^{\circ}$

A Monte Carlo simulation database was generated by using the CORSIKA air shower code \cite{corsika}. 
The simulations were carried out \cite{mkkoul} for the Mount Abu observatory altitude ($\sim 1300$m).
A total of 74000 $\gamma$-ray events in the energy range from 1-20 TeV were generated  
according to differential spectral index $2.6$. These events were generated  at five zenith angles:  $5^{\circ}$, $15^{\circ}$, $25^{\circ}$, $35^{\circ}$ and $45^{\circ}$. 
Because of the technical limitation of Monte Carlo simulation, zenith angles have discrete values whereas the observational hadrons have a continuous distribution in zenith 
angle. It is important to take this fact into account. The simulated training sample was prepablack in such a way that the zenith angle distribution as 
well as the size distribution for $\gamma$ and hadron showers were as close as possible to those of the real data. 
We approximated all the observational zenith angles between $0-10^{\circ}$ by  $5^{\circ}$, $10-20^{\circ}$ by  $15^{\circ}$ etc. 
The camera trigger was simulated, taking into account photon losses due to wavelength dependent photon absorption, reflection coefficient
of the mirror facets, light concentrators used in the camera and the quantum efficiency of photomultiplier tubes. All the triggeblack events 
underwent the usual image cleaning procedure \cite{cleaning} to eliminate the background noise.  In order to have a 
robust 
and well contained image inside the camera, prefiltering cuts of size (photoelectrons) 
$\geq 50$ and $0.4^{\circ} \leq$ distance $\leq 1.4^{\circ}$ were applied. It yielded a total of $60000$ events for $\gamma$-rays. It is 
not advisable to use Monte Carlo hadrons while using the 
observation data because the hadronic showers have large fluctuations. Moreover, the generation of hadron showers is much more time consuming owing 
to very small trigger probability. In the presence of ON or OFF source observational data, there is no need to use the simulated hadrons, the  
cosmic ray hadrons from the observation can be used. In the present study, the actual events recorded by the TACTIC telescope 
are used instead of simulated hadrons. 

\section{$\gamma$/hadron classification methods}
The problem of $\gamma$/hadron segregation is formulated as a two class problem: Viz $\gamma$ -ray showers represent one class, hadronic showers represent the second. 
Here we will apply the Random Forest method for $\gamma$/hadron classification. As mentioned earlier \cite{tactic2007}, the previous analysis of this data was carried out by using the 
Dynamic Supercuts method. Both methods are described below.

\subsection{Dynamic Supercuts method}
\label{sect:dsc}
In the conventional Hillas parameter technique, various sequential 
cuts in the image parameters are applied so as to maximize the $\gamma$-ray like signal and reject maximum number of background events. However, 
this scheme has a disadvantage because the width and length parameters  grow with the primary energy. It is observed that width and length of an image 
are well correlated with the logarithm of size. The size of the image provides an estimate of the primary energy. The process of scaling the width 
and length parameters with the size parameter is known as the Dynamic Supercuts method. By employing this method, the optimal number of cut 
parameters and their values were estimated by numerically maximizing (over the Hillas parameters) the so called \textquotedblleft quality factor$\textquotedblright$ \cite{gaug}, 
defined as 
\begin{equation}
\mathrm{Q} = \frac{\mathrm{\epsilon}_\mathrm{\gamma}}{\sqrt{\mathrm{\epsilon_P}}}
\end{equation}
Here $\epsilon_{\gamma}$ and $\epsilon_P$ are $\gamma$ and hadron acceptances respectively. The set of image parameters after quality factor maximization are shown in (Table 1).

\begin{table}[h]
\caption{ Dynamic Supercuts  selection  criteria \cite{tactic2007}}
\centering
\begin{tabular}{|c|c|}
\hline 
Parameter  & Cut Values\\
\hline
LENGTH (L) & $0.11^\circ\leq L \leq(0.235+0.0265 \times \ln S)^\circ$\\
\hline
WIDTH  (W) & $0.06^\circ \leq W \leq (0.085+0.0120 \times \ln S)^\circ$\\
\hline
DISTANCE (D) & $0.52^\circ\leq D \leq 1.27^\circ cos^{0.88}\theta$ ;($\theta$$\equiv$zenith ang.)\\
\hline
SIZE (S)  & $S \geq 450 d.c$ ;(6.5 digital counts$\equiv$1.0 pe )\\
\hline
ALPHA ($\alpha$) &  $\alpha \leq 18^\circ$\\
\hline
FRAC2 (F2) &  $F2 \geq 0.35$ \\
\hline
\end{tabular}
\end{table}



\subsection{Random Forest Method} 
This is a multivariate method based on the aggregation of a large number of decision trees. In this method, Bagging and Random Split Selection is combined 
to construct a common classifier. The Random Forest algorithm was developed by Leo Breiman and Adele 
Cutler \footnote{http://www.stat.berkeley.edu/~breiman/RandomForests/}. The decision trees are constructed by partitioning the dataset recursively. Tree 
construction begins with the root node containing the entire dataset and ends at the leaf node. Every node is then assigned a class. Finally a voting 
decides on the class assignment for each event. For a detailed description of the Random Forest method, the reader is referblack to 
our recent work \cite{mradul} as well as \cite{magic}. 

\section{$\gamma$/hadron discrimination using the Random Forest method}
\begin{figure}[!h]
\begin{center}
\includegraphics[width=0.35\textwidth,angle=270,clip]{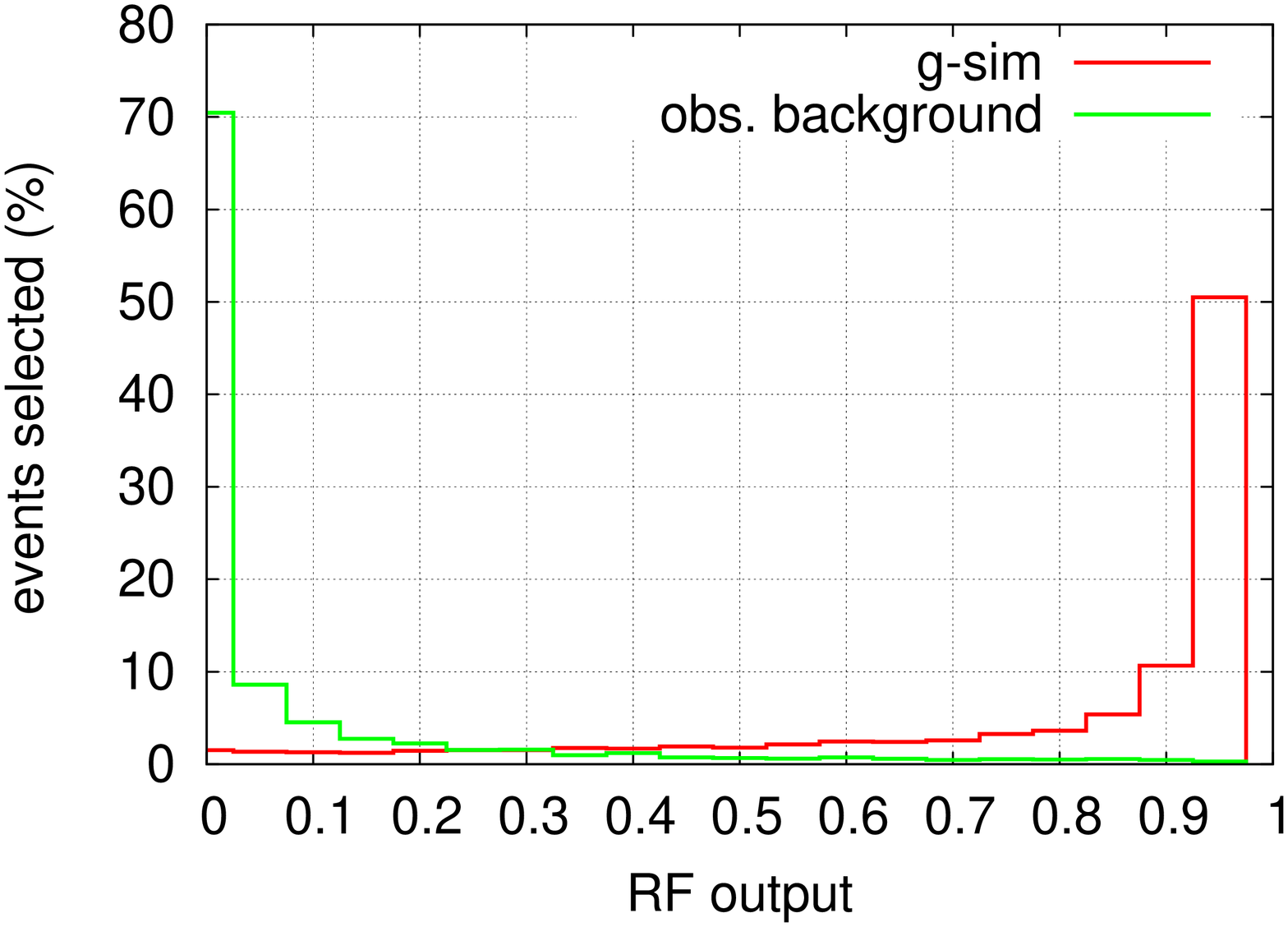}\includegraphics*[width=0.35\textwidth,angle=270,clip]{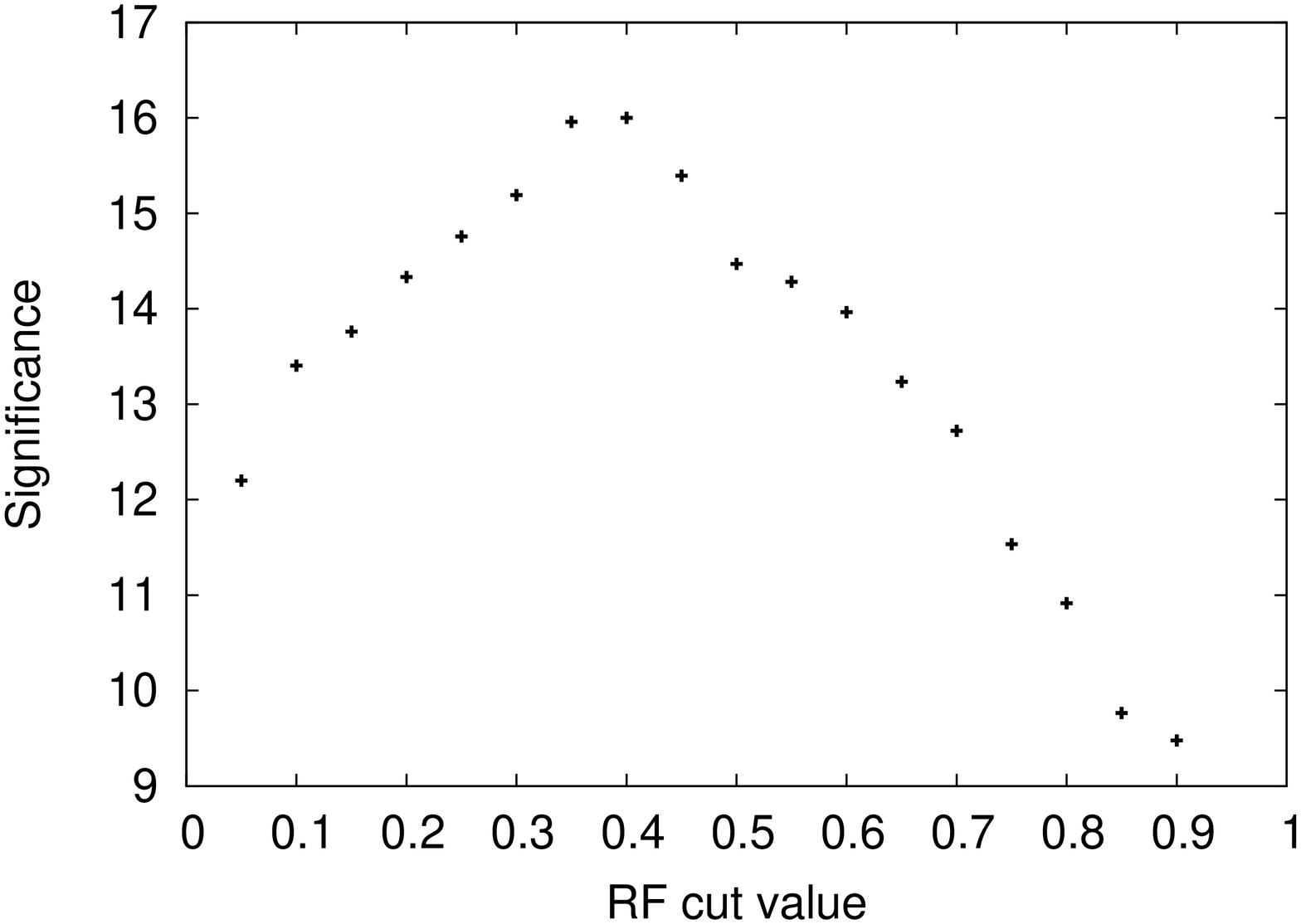}
\caption{\label{Figure.}The left panel shows the simulated $\gamma$ and observational background events for the estimation of optimum cut value of 
the Random Forest output. A total of 34120 events for simulated $\gamma$ and actual background events were employed as a test sample. 
The right panel shows the significance as a function of Random Forest cut value.}\label{Figure:class}
\end{center}
\end{figure}

A total of $60,000$ simulated $\gamma$-ray events were obtained from simulation after applying the precuts defined in the earlier section. 
In order to have a balanced data set for $\gamma$ and hadrons, we extracted a total of $60,000$ 
events from the ON source observations (with $\alpha \geq 27^{\circ}$) available from Mrk 421 and Crab Nebula observations carried out in $2005-06$. A total of 
$30,000$ events each from Mrk 421 and Crab Nebula observations were extracted respectively. The ON source events above this alpha range are pblackominantly background events. There is a 
possibility of a small gamma contamination above  $\alpha \geq 27^{\circ}$ but the Random Forest algorithm is sufficiently stable to deal with $\sim 10\%$ contamination. 
This was checked by successively contaminating a 
proton Monte Carlo  sample with Monte Carlo gammas. 
In total, $70\%$ events from the $\gamma$-sample and $70\%$ of the events from the hadron-sample were employed to train the Random forest and the remaining 
$30\%$ events were used as a test sample. 
The frequency distribution of pblackicted $\gamma$ and hadron events from Random Forest for a test sample containing 17820 $\gamma$  and $16300$ background events  
as a function of Random Forest cut value is shown in Figure \ref{Figure:class}. 
For determining the output cutoff value from the Random Forest method to optimize the  separation of the two event classes, we maximized   
the statistical  significance ($ N_{\sigma}$) of the signal following \cite{bock}. The statistical significance is defined as 
\begin{equation}
N_{\sigma}=\frac {N_{\gamma}} {\sqrt {  N_{\gamma}+ 2 N_p}}
\end{equation}
where  $N_{\gamma}$  and   $N_{p}$  are the  number of $\gamma$-rays and  hadrons, after classification, respectively. A cut value of $0.35$ was chosen on the basis of optimized significance. 
It is to be noted that the present cut value is obtained by carrying out the simulation for the Crab Nebula. Once the significance is optimized, the same cut value is used for Mrk421.

\section{Application of Random Forest method for Mrk 421 observation}
The recorded data with the telescope was subjected to the standard image cleaning 
procedure \cite{konopelko} by using the picture and boundary threshold of $6.5 \sigma$ and $3 \sigma$. All such events were processed to characterize the images using the Hillas parameters.  
Typically $\gamma$-ray events have narrow elliptical shapes whereas hadronic events are more irregular. These differences in image 
shape are used for segregating $\gamma$/ hadron events. 
The Dynamic Supercuts method yields an excess of $(1236 \pm 110)$ $\gamma$-ray events with a statistical significance of 11.5 $\sigma$ \cite{tactic2007}. 


\subsection{Alpha plot analysis for Mrk 421 using Random Forest method}
The $\alpha$ distribution determines the excess number of $\gamma$-ray events. The background events were extracted 
from $27^{\circ} \leq \alpha \leq 81^{\circ}$ whereas $\alpha \leq 18^{\circ}$ defines the region of signal events. Before employing the Random Forest for 
Mrk421 data, we validated the method by analyzing the Crab Nebula data observed during nearly the same lunation (November 10, 2005 to January 30, 2006).
The Random forest method yielded a total of ($1080 \pm 113$) events for a total of $\sim 101.44$ h of observation. The same data was analyzed by restricting the zenith 
angle of observation from 15-45$^{\circ}$, a zenith angle range similar to that of Mrk 421 observation. A total of $(634 \pm 73)$ events were obtained 
in an observation time of $\sim 63.33$ h with a corresponding $\gamma$ ray rate of $\sim (10.01 \pm 1.1$)  h$^{-1}$. This $\gamma$-ray rate is designated as 
a reference of 1 Crab Unit (CU).
The same trained forest was employed for estimating the number of excess events as well as the energy spectrum for Mrk 421. All the observation spells were analyzed individually. 
Table 2 shows the spell wise analysis.
\begin{table}[h]
\caption{Detailed  Spell wise  analysis report of  Mrk 421 data due to RF (DSC)}
\centering
\begin{tabular}{|c|c|c|c|c|c|c|}
\hline 
Spell & Obs. time  &$\gamma$-ray   & $\gamma$-ray rate  &  Significance  \\ 
      &  ( h. )  & events      &  (h$^{-1}$) &  ($\sigma$)    \\
\hline
 I    &    9.24   & 23 (9)     & 2.49 (0.97)     & 1.7 (0.4)   \\
\hline 
 II   &   35.71   & 322 (275)   & 9.01 (7.70)    & 6.9 (5.8)   \\
\hline
 III   &   61.53   &730 (676)  & 11.86  (10.99)   & 13.8 (10.6)    \\
\hline
 IV   &    34.54   & 131 (91)   & 3.79 (2.64)     & 3.2 (1.9)    \\ 
\hline
 V     &   31.14   & 100 (61)   &  3.38(1.96)    & 3.2 (1.6)      \\
\hline
 VI    &   29.55   & 151 (123)  & 5.10 (4.16)    & 4.0 (3.8)       \\
\hline
\hline
 All data &  201.72 & 1457 (1236)  & 7.22 (6.13)   & 14.6 (11.5)      \\
\hline
 II +III  &  97.24  & 1052 (951)   & 10.82 (9.78)  & 13.8 (12.00)      \\
\hline
\end{tabular}
\end{table}   
The significance of the signal was calculated by using Li \& Ma \cite{li1983}.

\begin{figure}[!h]
\begin{center}

\includegraphics[width=0.35\textwidth,angle=270,clip]{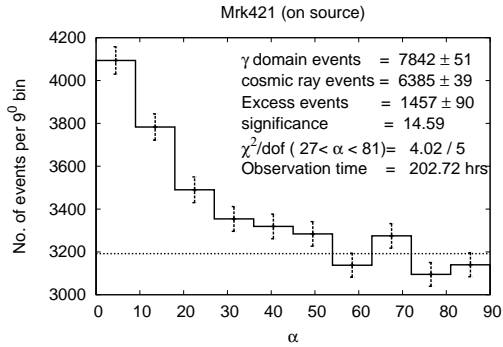}
\caption{\label{Figure.} On source $\alpha$ plot for Mrk 421 during December 07, 2005 - April 30, 2006 for $\sim 202$ h. The expected background events were 
obtained using the background region ($27^{\circ} \leq \alpha \leq 81^{\circ})$}\label{Figure:alp-mrk421}
\end{center}
\end{figure}
The data from all the spells (I-VI) yielded a total of $(1457 \pm 90)$ events with a statistical significance of 14.6 $\sigma$. Figure \ref{Figure:alp-mrk421} shows the $\alpha$ 
plot for the Mrk 421 source for the entire observation $\sim 202$ h. It demonstrates that compablack to the Dynamic Supercuts method, which produced a total of $(1236 \pm 110)$ 
events with a statistical significance of 11.5 $\sigma$, the Random Forest method leads to an improvement in the excess events by $\sim 18\%$ and significance by $\sim 26\%$.

\section{Energy spectrum of Mrk 421}
The primary $\gamma$-ray energy reconstruction was carried out by applying the Artificial Neural Network method (ANN) \cite{tactic2007}. 
The $\gamma$-ray energy reconstructed with a single imaging telescope, in general, is a function of image size, distance and zenith angle. 
The energy estimation was carried out by employing a 3:30:1 ( i.e.  3 nodes in the input layer, 30 nodes in hidden layer and 1 node in the output layer) 
configuration of ANN with a back propagation training  algorithm  \cite{rei1994}. We used a total of 10,000  $\gamma$-ray showers for training the network. 
The showers were generated at five zenith angles ($5^{\circ}$, $15^{\circ}$, $25^{\circ}$, $35^{\circ}$ and $45^{\circ}$). The effective area as a function of zenith angles 
and energy was estimated by the standard procedure  \cite{mohanty}. The trained network provided the weights which were employed to estimate the energy of 
excess $\gamma$-ray events.

The differential photon flux per energy bin is a direct function of zenith angle, energy-dependent effective area and $\gamma$-ray retention.  Eventually, the spectrum 
was obtained by using the formula 
\begin{equation}
\frac{d\Phi}{dE}(E_i)=\frac {\Delta N_i}{\Delta E_i \sum \limits_{j=1}^5 A_{i,j} \eta_{i,j} T_j}
\end{equation}
where $\Delta N_i$ and $d\Phi(E_i)/dE$ are the number of events and the differential  flux at energy $E_i$, measublack in the ith  energy bin $\Delta E_i$ and over the zenith angle range 
of 0$^\circ$-45$^\circ$ respectively. $T_j$ is the observation time in the jth zenith angle bin with corresponding energy-dependent effective area ($A_{i,j}$) and $\gamma$-ray acceptance ($\eta_{i,j}$). 
The 5 zenith angle bins (j=1-5) used are 0$^\circ$-10$^\circ$, 10$^\circ$-20$^\circ$, 20$^\circ$-30$^\circ$, 30$^\circ$-40$^\circ$  and 40$^\circ$-50$^\circ$ with  effective  collection area  and  
$\gamma$-ray acceptance  values   available at 5$^\circ$, 15$^\circ$, 25$^\circ$, 35$^\circ$ and 45$^\circ$. The number of $\gamma$-ray events  ($\Delta N_i$)  in a particular  energy bin is  
calculated  by subtracting the expected number of background events from the  $\gamma$-ray domain events. In order to validate the trained ANN, the spectrum of the 
Crab Nebula was reproduced. The spectrum of the Crab
nebula is fitted by a power law $(d\Phi/dE=f_0 E^{-\Gamma})$  with  $f_0$ $\sim (1.94 \pm0.15) \times 10^{-11}$ cm$^{-2}$ s$^{-1}$ TeV$^{-1}$ and $\Gamma \sim2.66\pm0.09$. 

\begin{figure}[!h]
\begin{center}
\includegraphics[width=0.30\textwidth,angle=270,clip]{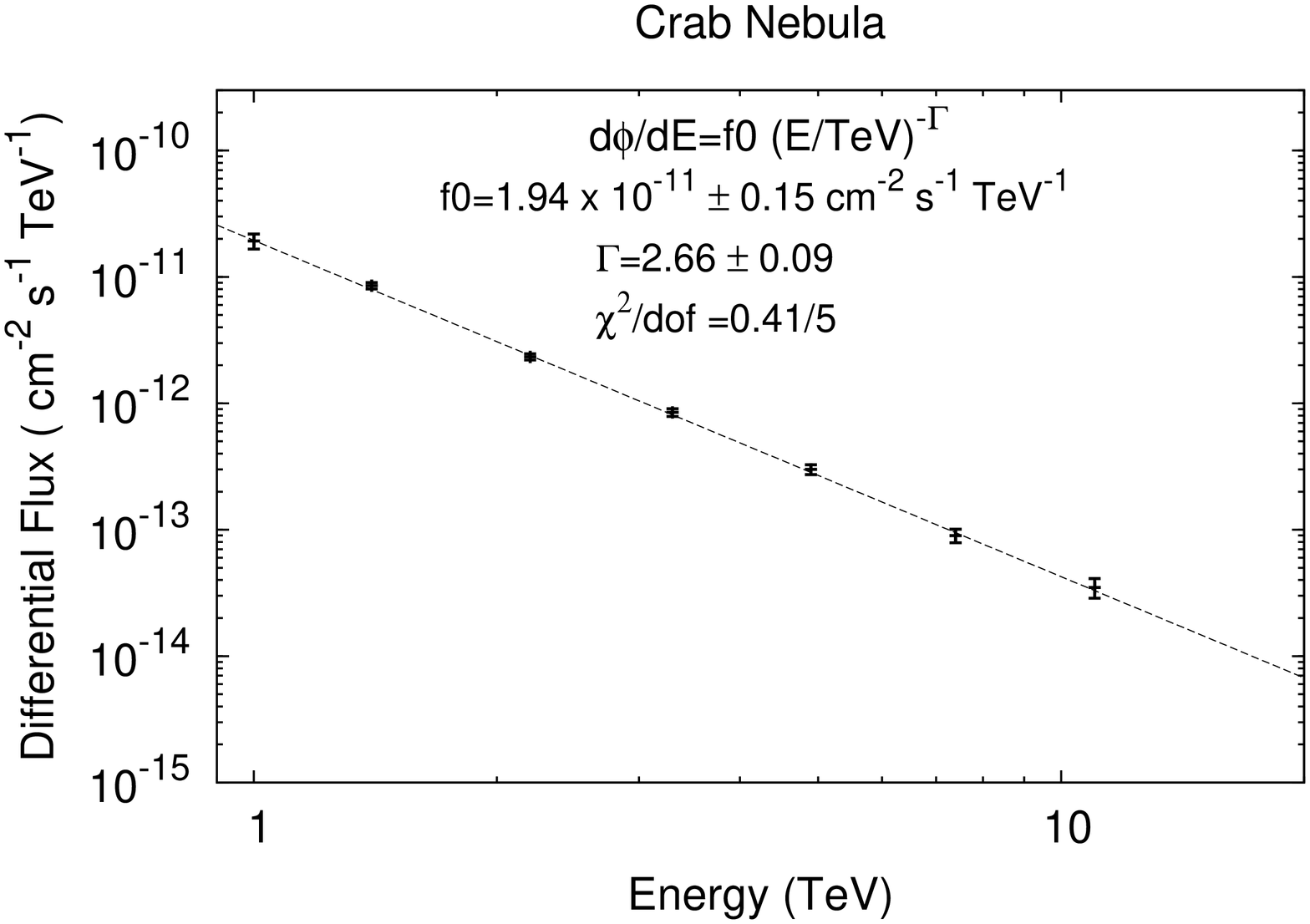}\includegraphics[width=0.30\textwidth,angle=270,clip]{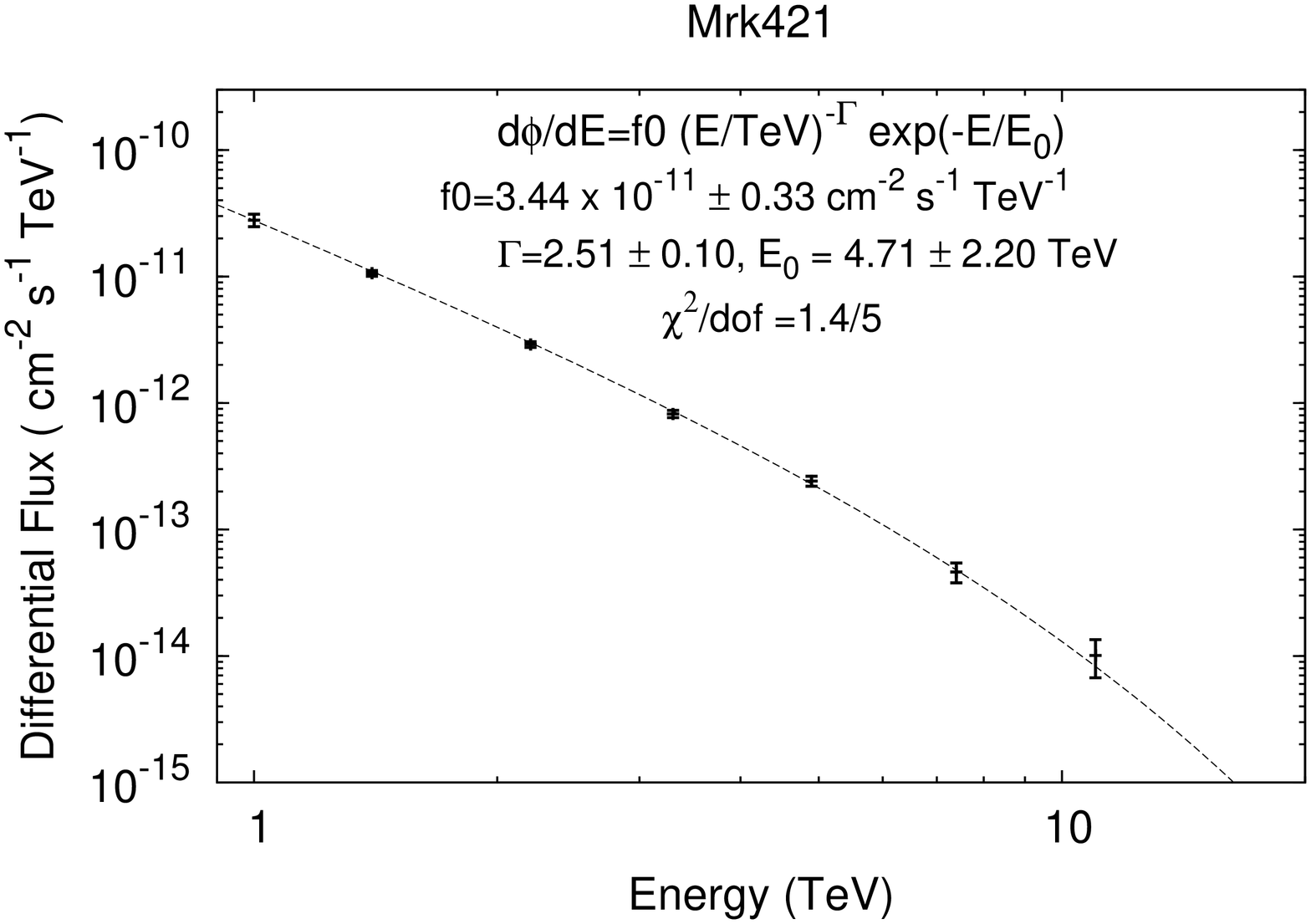}
\caption{\label{Figure.}Left panel shows the Crab Nebula spectrum during November 10, 2005 - January 30, 2006 for $\sim 101$ h. Right panel shows the Mrk 421 spectrum during Spell II and III during December 07, 2005 - April 30, 2006 for $\sim 97.24$ h. }\label{Figure:spec}
\end{center}
\end{figure}

After reproducing the Crab Nebula spectrum, we obtained the spectrum of Mrk 421 for the spells II and III where the source was emitting at more than 1 CU.  
The resultant spectrum is well fitted by a simple power law with exponential cut off $(d\Phi/dE=f_0 (E/E_{0})^{-\Gamma})$  with  $f_0 \sim (3.44 \pm0.33) \times$  10$^{-11}$ cm$^{-2}$ s$^{-1}$ TeV$^{-1}$  
and $\Gamma \sim2.51\pm0.01$ and $E_{0} = 4.71 \pm 2.20$. Both the obtained spectra (Figure \ref{Figure:spec}) match well with the previously reported results \cite{tactic2007}.


\section{Results and Discussion}
\begin{figure}[!h]
\begin{center}
\includegraphics[width=0.30\textwidth,angle=270,clip]{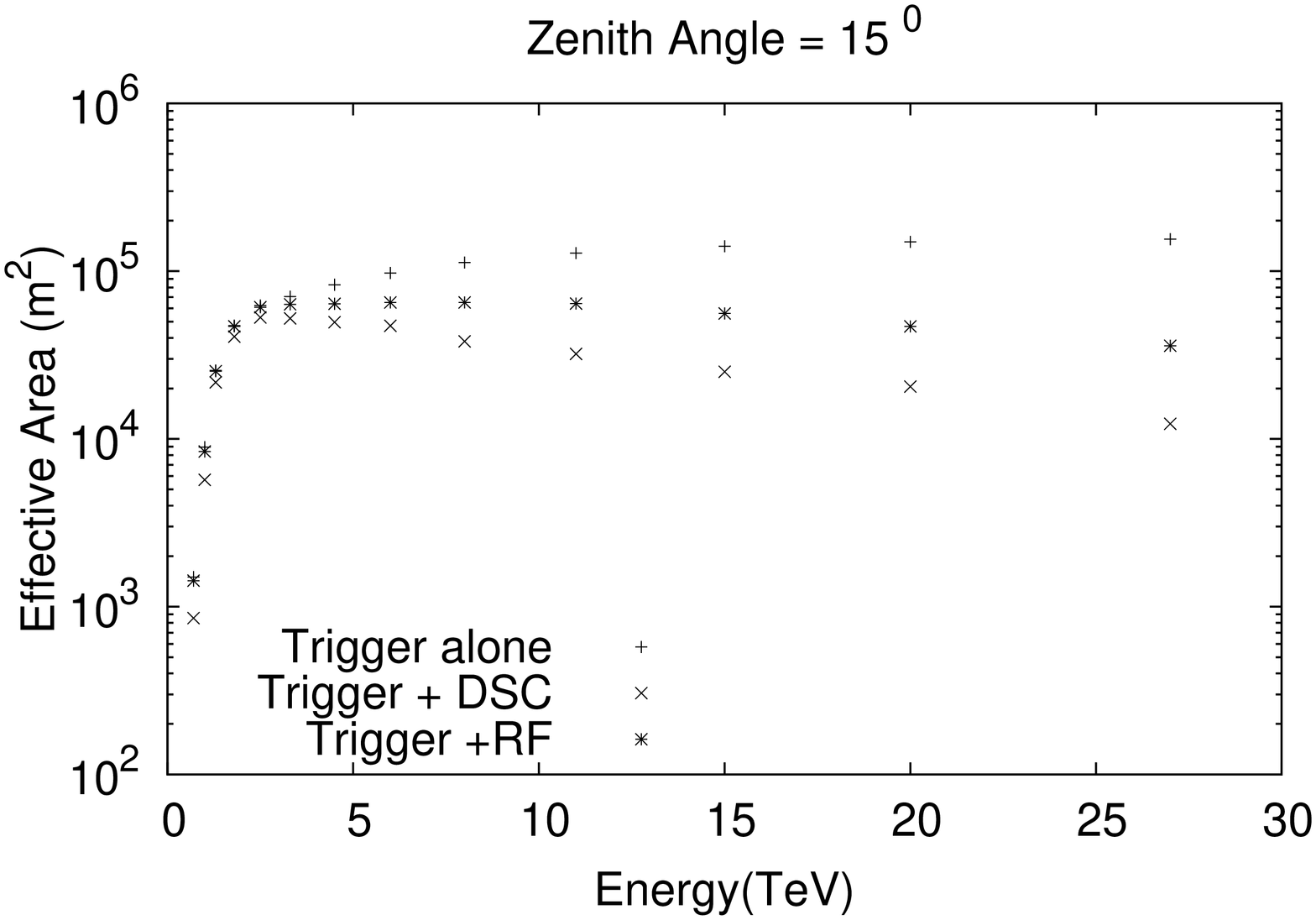}\includegraphics[width=0.30\textwidth,angle=270,clip]{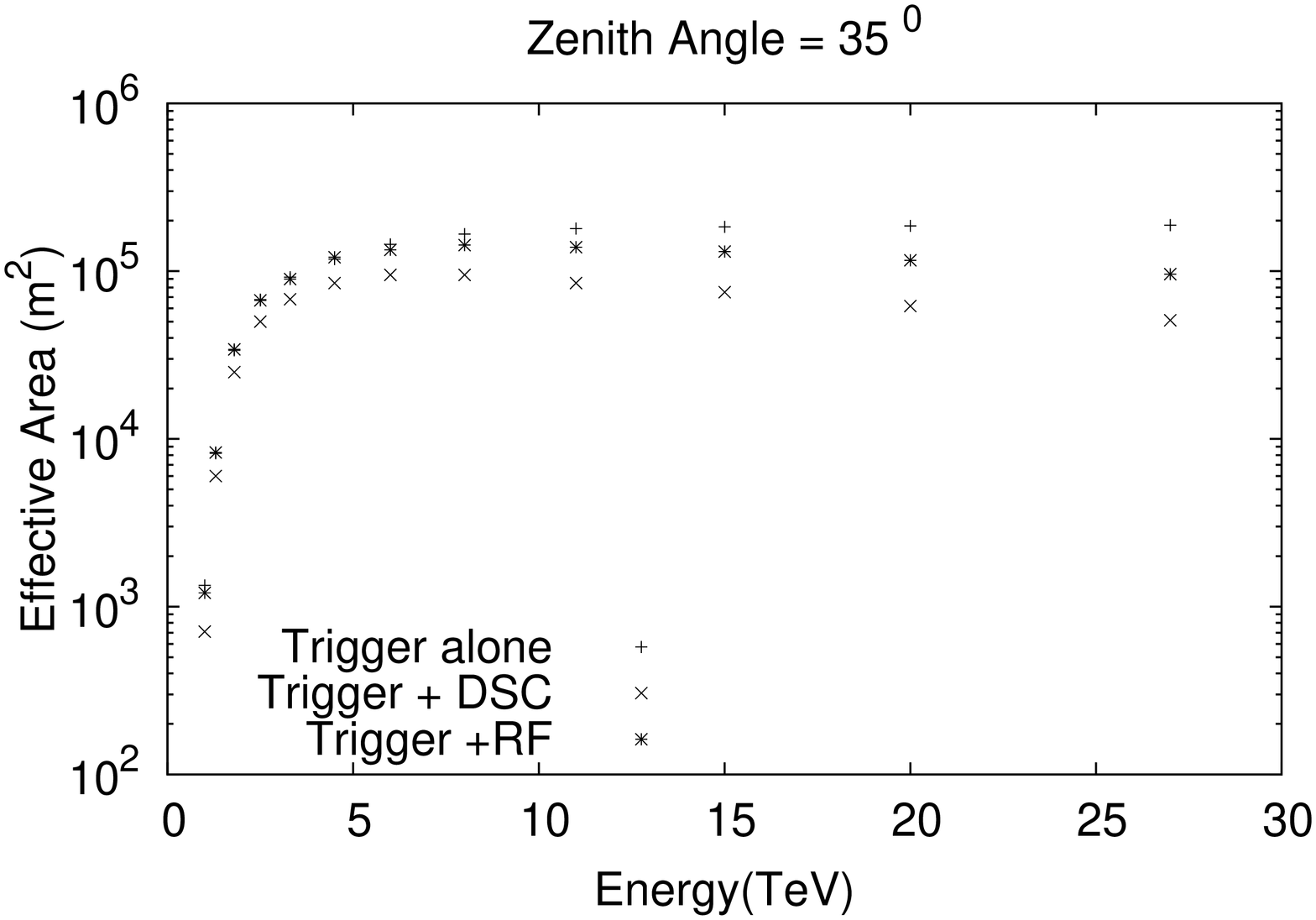}
\caption{\label{Figure.}Effective collection area as a function of $\gamma$-ray energy at two zenith angles ($15^0, 35^0$). The top most curve represents the effective 
collection area for the events triggering the telescope. Two other curves represent effective collection area after the application of Dynamic Supercuts and Random Forest method.}\label{Figure:effarea}
\end{center}
\end{figure}

The extragalactic source Mrk 421 was observed by the TACTIC imaging telescope  during December 
$07, 2005$ - April  $30, 2006$ for $\sim 202$ h. Previous analysis of this data using the Dynamic Supercuts method led to a detection of flaring activity from the source 
at Energy $ > 1$TeV. Here we re-analyzed this data by using the machine learning Random Forest method. 
The Dynamic Supercuts method as well as the Random Forest method showed that Mrk 421 was in the high state in  Spell II and III. The key result of this study is that the 
Random Forest method estimated more excess $\gamma$-ray like events 
compablack to  previous results \cite{tactic2007}. This is so because  Supercuts method yielded a total of $(1236 \pm 110)$  excess events for the entire 
data set, while the Random Forest method resulted in ($1457 \pm 90)$ excess events. The signal strength estimated by the Random Forest method was 14.6 $\sigma$ as compablack 
to 11.5 $\sigma$ obtained by the Dynamic Supercuts method. The energy spectrum of Mrk 421 as measublack by the TACTIC telescope for spell II and III is compatible with a power law with exponential cut off.
The Random Forest method yielded $\sim 18\%$ more events than the Dynamic Supercuts method. In order to find out the energy dependance of excess $\gamma$-ray events, we compablack the energy 
dependent effective area at two representative zenith angles ($15^0, 35^0$). Figure \ref{Figure:effarea} shows the effective collection areas of the telescope after the application of Dynamic Supercuts and 
Random Forest method as a function of $\gamma$-ray energy.
It is clear that the application of the Random Forest method leads to higher effective area beyond $E_{\gamma} = 9$ TeV, i.e. the Random Forest method is able to collect more high energy events compablack to the 
Dynamic Supercuts method. 
In addition to the source detection in high state in Spell II and III, TACTIC also observed this source during 
the last lag of Spell VI in the month of April  2006. It may be mentioned that during this period, this source was also observed by various groups. The Whipple telescope \cite{horan2009} carried out 
the observation in April and June 2006. The MAGIC telescope observed this source from April 22 to April 30, 2006 \cite{magic2010}. Both experiments detected 
the source in high state. However, TACTIC has only a 4.1 $\sigma$ detection during  this period which might be owing to its lesser sensitivity.
It should be noted that the TACTIC readout employs charge integration over the full gating window. It contrasts to MAGIC where FADC-readout \cite{readout} is used, 
allowing to exploit the time-development of the image for gamma/hadron separation. It has been demonstrated that the use of timing information in the analysis of the MAGIC data 
results in an enhancement of about a factor 1.4 of the flux sensitivity to point-like sources. The sensitivity of the TACTIC telescope might improve by adopting this strategy.

The applicability and superiority of the Random Forest method over the Dynamic Supercuts method for the TACTIC telescope has been demonstrated by the present study. 
In a subsequent work, this method will be employed for available data of various sources corresponding to both detection and non-detection by the Dynamic Supercuts method.

\section*{Acknowledgements}
MS thanks P. Savicky for making available the decision plot of the simulated MAGIC data\footnote{\small http://archive.ics.uci.edu/ml/datasets/MAGIC+Gamma+Telescope/}, N. Bhatt, K.K.Yadav for 
useful discussions, H. Bhatt and B. Ghosal for re estimating the excess $\gamma$-ray events using the Dynamic Supercuts method. The authors also thank all the colleagues of the 
instrumentation and observation team.

\bibliographystyle{h}
\bibliography{mradul-nimpa}

\end{document}